# Fog Computing Vs. Cloud Computing

**Moonmoon Chakraborty**

Oracle and Cloud (AWS) database Engineer and DevOps Architect

Master of Science in Information Technology

University of the Cumberlands

Chicago, United States

## Abstract

This article gives an overview of what Fog computing is, it's uses and the comparison between Fog computing and Cloud computing. Cloud is performing well in today's World and boosting the ability to use the internet more than ever. Cloud computing gradually developed a method to use the benefits of it in most of the organizations. Fog computing can be apparent both in big data structures and large cloud systems, making reference to the growing complications in retrieving the data accurately. Fog computing is outspreading cloud computing by transporting computation on the advantage of network systems such as cell phone devices or fixed nodes with in-built data storage. Fog provides important points of improved abilities, strong security controls, and processes, establish data transmission capabilities carefully and in a flexible manner. This paper gives an overview of the connections and attributes for both Fog computing and cloud varies by outline, preparation, directions, and strategies for associations and clients. This also explains how Fog computing is flexible and provide better service for data processing by overwhelming low network bandwidth instead of moving whole data to the cloud platform.

**Literature Review or Background**

Fog computing is a distributed computing structure where data logically stored in a location between the data source and the cloud, also known as fog networking. Fog networking is closely associated with cloud computing. The benefits of Fog computing over the cloud are huge as the trends are now changed and all companies are showing their interests in the technical innovation. Configuring Fog computing provides more choices for processing data in a suitable way which benefits the organizations. Fog computing is different than cloud computing. A cloud is a centralized system which describes many users to be available over the internet, while fog is a distributed computing structure which works as an intermediary device between remote servers and the hardware. Fog computing has control on the data which can be sent to the server and which can be managed locally.

The company Cisco have invented the tern fog computing (or fogging) in 2014, so it's almost new for people. Fog and cloud are interrelated, and fog is closer to the Earth than cloud, in technology also the same steps are followed. Fog is very closed to end users brings the cloud capabilities down to the ground to get the flexibility, productivity, scalability which are the benefits of the cloud. Fog consists of multiple edge nodes which can connect to physical devices directly as shown in the below picture. The fog edge nodes are closer to the physical devices and that is the reason why fog is capable to deliver instance connections. The significant processing power of edge nodes permits to achieve computation of a huge quantity of data on their own, without transferring it to servers resided in a far distance.

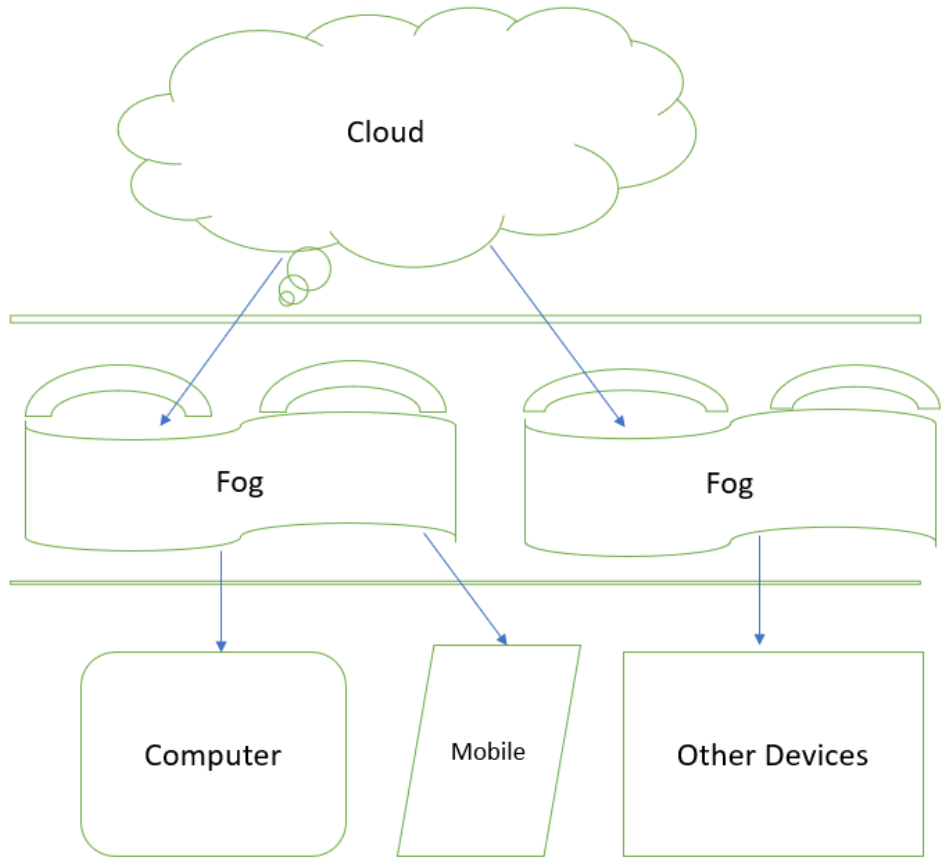

**Fig1: Fog connected to cloud and physical devices**

**Discussion**

Cloud computing is developed to use the remote servers or computers across the internet to accomplish operations, data storage and computing power instead of using a local computer or server. Cloud computing allows the service to be delivered over the internet. These services include storage, network, applications, data etc.

Fog computing is a term coined by Cisco and Fog is the extension of services beyond cloud computing. Fog computing contains a decentralized environment for computing in which the infrastructure provides data, computations, applications, and storage. Fog computing doesn't access the remote computers, in spite, it uses the local computers the purpose of computation producing a reduction of latency issues and performance additionally making it more powerful and well-organized.

Let's discuss the advantages and disadvantages of Fog computing also the comparison of cloud and Fog computing.

**Advantages:**

Security: Since fog computing is connected to various nodes and the system is very complex so it maintains high security. Fog computing has all applicable security controls and processes in place which solves the problems of security risks and mitigate the threads.

Great bandwidth: In Fog computing the pieces of information are circulated via different channels during different times which is a huge example of great bandwidth systems.

Less latency: Latency is the delay between the transfer from one point to another. As Fog is closed to the end-users, the delay during the transfer is ignorable. No connection loss:

Power efficiency: Power efficiency is one of an important advantage. Many likes to use Fog computing due to its quality of being efficient, faster performance and processes.

Great user experience: Since there is no downtime, users like to use Fog computing. Users get an immediate response as soon as data is moved to the cloud, big data or real-time analytics.

**Disadvantages:**

Expensive: Fog computing is very expensive because the organization needs to buy devices like hubs, routers, gateways.

Complex system: Fog computing uses many nodes and it is the additional layer in the data processing and storage systems, so it is a very complicated computing process.

Scalability: Cloud is more scalable than Fog computing.

**Difference between cloud and Fog computing:**

1. Cloud is centralized and contains a large amount of data that can be positioned around the globe, far away from client devices. Fog is scattered and contains several small nodes situated close to the client devices.

2. Fog is the layer between the cloud and the devices like computer, laptop, mobile etc. As Fog acts as a mediator, it is less time consuming to transfer the data. When there is no layer then cloud needs to communicate directly to the end devices which take more time than using Fog computing.

3. Cloud computing has low latency but not compare to Fog computing. Fog computing has low latency in terms of network.

4. Cloud computing doesn't provide any reduction in data while transferring data, but fog computing reduces data while sending to cloud.

5. Cloud computing preserves less bandwidth compare Fog.

6. In cloud computing, the response time of the system is low than Fog computing.

7. Cloud computing is secure, but Fog computing is highly secure. Since Fog is distributed and has complex architecture so Fog is secure than cloud.

8. In cloud computing only, multiple data sources can be integrated wherein Fog computing data source and the device can be integrated.

9. Cloud fails without internet but since Fog uses numerous procedures and standards, it doesn't fail if no internet connection.

10. Fig has a flexible infrastructure where the cloud has three models like PaaS (Platform as a Service), IaaS (Infrastructure as a Service) and SaaS (Software as a Service).

11. Fig computing has centralized supporting of user management where the cloud can be centralized or can be delegated to the third party.

12. Resource management is centralized in Fog computing wherein cloud it is centralized or distributed.

13. Future of cloud is Fog computing.

| 1. Parameters | Cloud Computing | Fog Computing |
|---|---|---|
| Goal | Immensity and powerful provisioning of IT administrations | Improve proficiency and performance of the process that is transported to the cloud for handling, examination, and storage |
| Computational focuses | Data requests are handled in the Cloud | Fog functions on the network edge |
| Abstraction Level | High | High |
| scalability Degree | High | High |
| Support of Multitask | Yes | Yes |
| Level Transparency | High | High |
| Run time | Real-time services | Real-time services |
| Type of Requests | Many small allocations | Many high allocations |
| Allocation unit | All shapes and sizes (wide & narrow) | All shapes and sizes (wide & narrow) |
| Level of Virtualization | Vital | Vital |
| Accessible type | IP | IP |
| Transmission | Device to Cloud | Device to Device |
| Security | Undefined | Possible, Determined |
| Infrastructure | 3 models (PaaS, IaaS, SaaS) | Flexible |
| Support of Operating System | A hypervisor (VM) on which multiple Oss can run | Hypervisor virtualization |
| Ownership | Single | Multiple |
| Service negotiation | SLA based | SLA based |
| Support of User management | Centralized or can be delegated to third party | Centralized |
| Resource management | Centralized/Distributed | Centralized |
| Allocation/Scheduling | decentralized/centralized | Centralized |
| Interoperability | Web Services (SOAP and REST) | Interoperability between heterogeneous resources. |
| Failure management | Strong (VMs can be easily migrated from one node to other) | Rescheduling of failed tasks |
| Service price | Utility pricing discounted for larger customers | Utility pricing and payment is made based on the uses |
| Type of service | IaaS, PaaS, SaaS, Everything as a service | CPU, network, memory, bandwidth, device, storage |
| Example of real world | Amazon Web Service (AWS), Google apps | Significant Fog applications involves real-time interactions rather than 228 batch processing. |
| Response Time | High | Low |
| Critical object | Service | Service |
| Number of users | Unlimited | Unlimited |
| Resource | Unlimited | Unlimited |
| Future | Fog Computing | Next Generation of Internet and computing |

**Table: Fog vs Cloud Computing**

**Conclusion**: This article explains the comparison between Fog and cloud computing. Cloud computing technology now established, and many growth expansions exists for design and implement cloud infrastructure. Fog computing is an initial phase of research and still, example replicas and development tools are under research phase, but this is understood that future of Fog computing in modern computing technology will evolve quickly and utilized edge of devices for computational resources.

**AUTHOR'S PROFILE**

Moonmoon Chakraborty has more than 10 years of experience in professional services and dynamic consultants on the overall design, configuration, delivery of software, data solutions to meet customer requirements. She is also engaged in Data Management practice with deep focus on Data Architecting, Data Modeling, Data Integration, Data Conversion and DATA Quality. She is instrumental in diverse databases platforms such as Oracle, Teradata, PostgreSQL, DevOps and in AWS Cloud. Her passion is to interact with business teams, oversight and direction to project and technical team members in all stages of the implementation lifecycle and translating client desires into a solution.  Moonmoon is based out of Chicago. She holds bachelor's degree in Computer Sc & Engineering and Master of Science in Information Technology. She is certified with many Oracle database certifications, Prince2 foundation and practitioner, Agile Scrum Master, ITIL foundation and expert. She takes many initiatives at work, establishing strategic objectives, project plans, milestone and make dramatic improvement in business while meeting the project objectives and deadlines**.**